\documentclass[twocolumn,secnumarabic,amssymb, nobibnotes, aps, prd]{revtex4-1}

\usepackage{graphicx}
\usepackage{amsmath}
\usepackage{color}

\newcommand{\mb}{\mathbf}

\begin{document}

\title{Classical Imaging with Undetected Light}%

\author{A. C. Cardoso\textsuperscript{1}}%
\author{L. P. Berruezo\textsuperscript{1}}%
\author{D. F. \'Avila\textsuperscript{1}}%
\author{G. B. Lemos\textsuperscript{2}}%
\author{W. M. Pimenta\textsuperscript{3}}%
\author{C. H. Monken\textsuperscript{1}}%
\author{P. L. Saldanha\textsuperscript{1}}%
\author{S. P\'adua\textsuperscript{1}}%
\email{spadua@fisica.ufmg.br}

\affiliation{\textsuperscript{1}Departamento de F\'{\i}sica, Universidade Federal de Minas Gerais, Caixa Postal 702, Belo Horizonte, MG 31270-901, Brazil \\ \textsuperscript{2}International Institute of Physics, Federal University of Rio Grande do Norte, Campus Universit\'ario - Lagoa Nova, CP.1613 Natal, Rio Grande do Norte 59068-970, Brazil  \\ \textsuperscript{3}Instituto de F\'{\i}sica, Universidad Aut\'onoma de San Luis Potos\'{\i}, San Luis Potos\'{\i} 78290, M\'exico}

\begin{abstract}
We obtained the phase and intensity images of an object by detecting classical light which never interacted with it.  With a double passage of a pump and a signal laser beams through a nonlinear crystal, we observe interference between the two idler beams produced by stimulated parametric down conversion. The object is placed in the amplified signal beam after its first passage through the crystal, and the image is observed in the interference of the generated idler beams.  High contrast images can be obtained even for objects with small transmittance coefficient due to the geometry of the interferometer and to the stimulated parametric emission. Like its quantum counterpart, this three-colour imaging concept can be useful when the object must be probed with light at a wavelength for which detectors are not available.
\end{abstract}

\maketitle

The image obtained from an object illuminated by correlated photons has been studied in different experimental schemes usually referred to as quantum imaging \citep{2040-8986-18-7-073002,1464-4266-4-3-372}.
Sub-Rayleigh imaging \citep{PhysRevA.77.043832,PhysRevA.67.033812} and single photon imaging with structured light \citep{2040-8986-19-1-013001} are examples of applications of quantum light imaging. In a recent work, Lemos \textit{et al.} constructed the intensity and phase images of an object by detecting photons which have never interacted with it \citep{Nature}.  The authors used the pairs of entangled photons generated by spontaneous parametric downconversion as a quantum source, such that one of the photons could interact with the object while the other one was detected. The quantum correlations between the photons and a fundamental path indistinguishability in an interferometer \citep{PhysRevA.44.4614,PhysRevLett.67.318} were responsible for that result. The effect is distinct from the previous methods based on ghost images \citep{PhysRevA.53.2804,1367-2630-15-7-073032}, direct image  \citep{PhysRevLett.93.213903} or interaction-free imaging \citep{PhysRevA.58.605}, since it is necessary to detect only one photon from each pair,  the other one being discarded \citep{Advoptphoton.2.405}. The experiment reported in \citep{Nature} was theoretically explained in \citep{PhysRevA.92.013832}. A variation of the experiment of quantum imaging with undetected light, using three crystals, is analyzed by S. Ataman in a Gedankenexperiment demonstrating that an object can be undetectable \citep{Ataman2016}. The visibility of the interference pattern is responsible for the imaging contrast and can be optimized in different parametric down-conversion gain regimes as discussed theoretically by Kolobov \textit{et al.} \citep{2040-8986-19-5-054003}.

\begin{figure*}[htpb]
\centering
\includegraphics[scale=0.17]{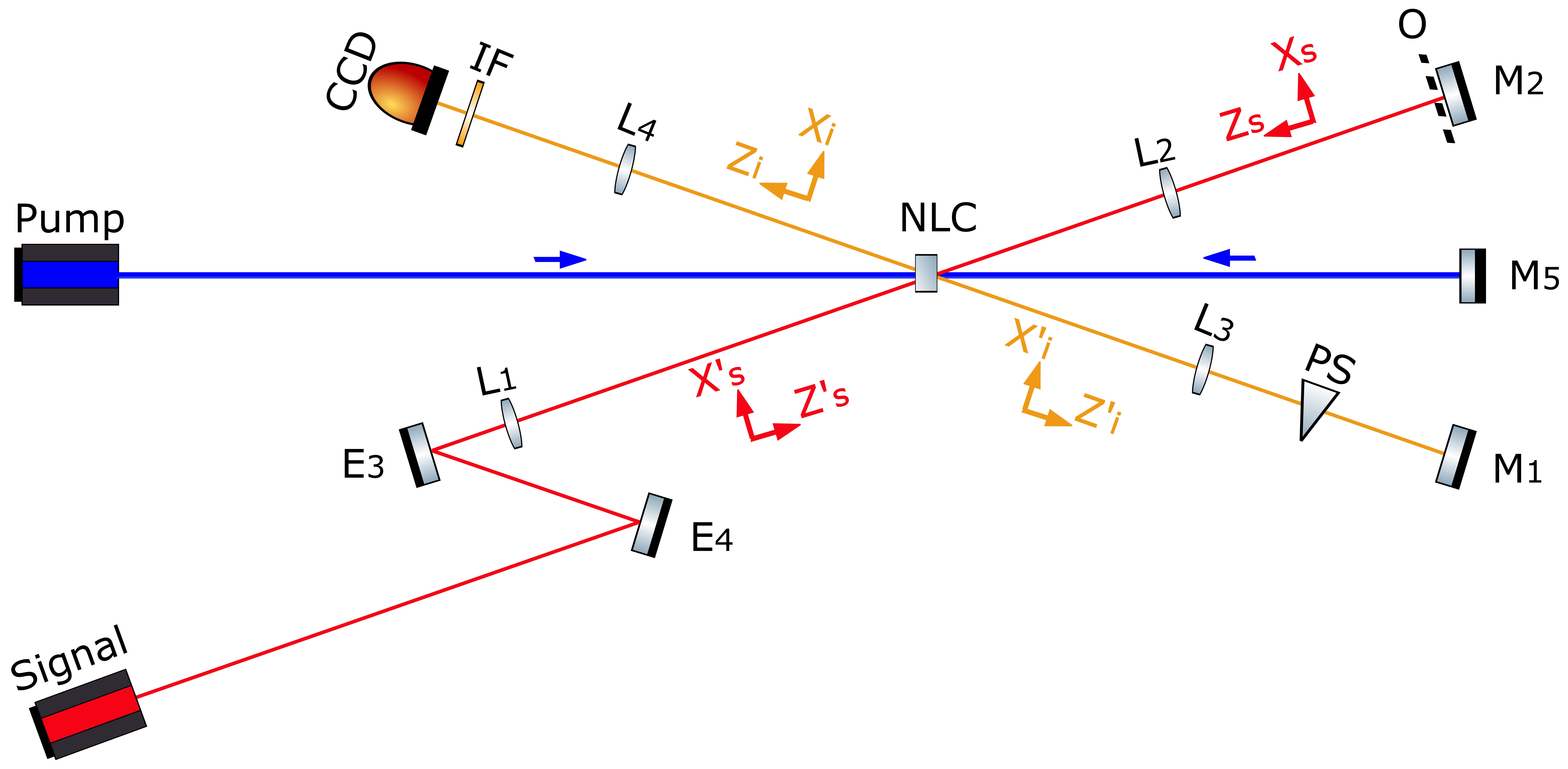}
\caption{Experimental setup. Type-II stimulated parametric downconversion is produced at a BBO nonlinear crystal NLC with the incidence of a vertically polarized pump beam and a vertically polarized signal beam. The pump beam is produced by a single mode laser diode with wavelength $405$\,nm and pumps the NLC with $7.2$\,mW power. The signal beam is produced by a laser diode with $830$\,nm and reaches the NLC with $2.8$\,mW power. The lens $L_1$, with focal distance $500$\,mm, focuses the signal beam on the crystal. A horizontally polarized idler beam is thus produced with wavelength  $792$\,nm. The angle between the pump beam and the signal (idler) beam is arround $9$ degrees ($8$ degrees). The pump beam is reflected backward by mirror $M_5$ for a second stimulated parametric downconversion process. The lens $L_2$, with focal length $150$\,mm, projects the Fourier transform of the signal field at the crystal on the object positioned very close to mirror $M_2$ and then the Fourier transform of the field reflected by $M_2$ with two passages through the object on the crystal again. The distances between NLC and $L_2$ and between $L_2$ and $M_2$ are $150$\,mm. The lens $L_3$, with focal length $150$\,mm, projects the Fourier transform of the idler beam at the crystal on the mirror $M_1$ and then the Fourier transform of the field at the mirror on the crystal again. The distances between NLC and $L_3$ and between $L_3$ and $M_1$ are $150$\,mm. The phase shifter $PS$ controls the relative phases between the idler beams generated in the first and second passages of the pump and signal beams through the nonlinear crystal. The lens $L_4$, with focal length $300$\,mm, projects the Fourier transform of the idler field at the crystal on the CCD camera. The distances between NLC and $L_4$ and between $L_4$ and the CCD are $300$\,mm. The bandpass interference filter (IF) is centered in $792$\,nm with bandwidth of $20$\,nm. The CCD camera used is a ICCD. Pixel dimensions of the CCD is 6.7 $\mu$m $ \times$ 6.7 $\mu$m. } \label{fig1}
\end{figure*}

In this work we show a classical analogue of the experiment by Lemos \textit{et al.} \citep{Nature}. In our experiment depicted in Fig. ~\ref{fig1}, a pump and a signal laser beams are injected into a non-linear crystal in the forward and in the backward directions, producing classical light (which we call “ idler”) through the process of stimulated down conversion \citep{CCPDIL}. An object is placed in the path of the amplified signal beam, after its first passage through the crystal. It is known that amplitude and phase distributions of the stimulating signal beam are transferred to the resulting idler beam \citep{OSAWang,PhysRevA.51.1631,PhysRevA.60.5074}. Our technique also relies on the classical version of the frustrated two-photon creation by interference \citep{PRL72}. The idler beams produced in the first and second passages of the pump and signal beams are perfectly aligned at the crystal and interfere, revealing the phase and intensity image  of the object, even though the idler beams did not interact with it. In this sense, we show that what is necessary for obtaining the object image by using light that did not interact with it are the spatial and phase correlations between the beam that interacted with the object and the detected beam, not necessarily entanglement as in \citep{Nature}.
A difference between our experiment and its quantum counterpart \citep{Nature} is that interference visibility is not a linear function of the object transmittance \citep{Pra270.05}, meaning that very high contrast images can be obtained even of objects that are nearly opaque \citep{2040-8986-19-5-054003}. Ref. \citep{CIWUP} predicts that, for objects with small mean transmittance,  the signal to noise ratio in a similar arrangement using a classical phase-conjugator source is higher than that which can be obtained in quantum imaging with undetected photons \citep{Nature}.
As in \citep{Nature} we can choose the illuminating wavelength and the detection wavelength independently, which is useful in applications where illumination must be at either end of the wavelength spectrum for which cameras are not easily obtained. 
Our setup is a particular instance of a non-degenerate nonlinear interferometer first discussed in  \citep{yurke} and shown to give an improved signal-to-noise ratio in phase measurements compared with a conventional interferometer \citep{plick,hudelist}. In our variant of the non-linear interferometer, the seed beam enters in the mode where an object is placed, and the other mode is used for imaging.

A simplified scalar theory is sufficient to explain the imaging phenomena, with the beams in the paraxial regime. The pump beam is considered to be a plane wave with angular frequency $\omega_{p}$ on both passages through the crystal. We assume that the signal beam is monochromatic with an angular spectrum given by $\tilde{A}(q_x', q_y'; \omega_{s})$ in its first passage through the nonlinear crystal.

An idler beam is generated in the first passage of the pump and signal beams through the nonlinear crystal \citep{boyd}. In Fig. ~\ref{fig1}, we represent the coordinate axes of the signal and idler beams that propagate in the $\mb{\hat{z}}_s'$ and $\mb{\hat{z}}_i'$ directions, respectively. With the reflection of these beams by the mirrors $M_1$ and $M_2$, we change the coordinate axes such that the beams propagate in the $\mb{\hat{z}}_s$ and $\mb{\hat{z}}_i$ directions. The phase matching conditions imply that  the incidence of the pump beam in a plane wave mode  and a signal beam with an angular spectrum $\tilde{A}(q_x', q_y'; \omega_{s})$ at the nonlinear crystal produce an idler beam with angular spectrum $\tilde{B}_1(q_x', q_y'; \omega_{i})=\alpha\tilde{A}(-q_x', -q_y'; \omega_{i})$ and an angular frequency $\omega_{i}=\omega_{p}-\omega_{s}$ , where $\alpha$ is a constant that depends on the nonlinear properties of the crystal and on the amplitude of the pump and signal beams \citep{boyd}. The intensity of the signal beam increases by a factor $1+|\alpha|^2$ and the intensity of the pump beam decreases by a factor $1-|\alpha|^2$, such that energy is conserved in the process.  In our experiment, we have $|\alpha|\ll1$, such that we can disregard the variation of the pump and signal beams intensities.

The lens $L_2$ projects the Fourier transform of the signal field at the crystal on the object plane positioned very close to the mirror $M_2$ \citep{goodman}. The field amplitude just before the object can thus be written as $E(x',y';\omega_s)=\mathcal{F}_{L_2}\left[\tilde{A}(q_x', q_y'; \omega_{s}) \right]$, with $\mathcal{F}_{L_2}$ meaning Fourier transform. The index $L_2$ is indicating that the scale of the Fourier transform depends on the focal length of the lens $L_2$. We assume that the object has a transmission coefficient $t(x,y)\mathrm{e}^{i\gamma(x,y)}$, with $t(x,y)$ and $\gamma(x,y)$ being real functions, in the coordinate axes given by $(x_s, y_s, z_s)$ in Fig. 1. The signal field becomes $E(x,-y;\omega_s)[t(x,y)]^2\mathrm{e}^{2i\gamma(x,y)}$ with the signal double passage through the object and its reflection by the mirror $M_2$. The inversion of the $y$ component of the field is associated to the inversion of the $y_s$ axis in relation to the $y_s'$ axis due to reflection. The lens $L_2$ then projects the Fourier transform of this field onto the nonlinear crystal.

In the second passage of the pump and signal fields through the nonlinear crystal, the idler field generated on the first passage is also present, but with a much smaller intensity than those fields, which should not result in extra nonlinear effects. 
The stimulation of the reflected idler beam in the second parametric down conversion was neglected in the calculation, such that the total idler field considered here is the sum of the idler fields in each passage with no amplification due to the reflected idler beam.
The second passage produces an idler field with angular spectrum $\tilde{B}_2(q_x, q_y; \omega_{i})=\alpha\mathcal{F}_{L_2}\left[ E(-x,y;\omega_i)[t(-x,-y)]^2\mathrm{e}^{2i\gamma(-x,-y)} \right]$. Note that the inversion of the signs of the $x$ and $y$ coordinates of the idler field in relation to the signal field is due to the fact that the phase matching conditions impose that a transverse wavevector component $\mb{q}$ in the  $x_sy_s$ plane of the signal field produces an idler beam with transverse wavevector component $-\mb{q}$ in the $x_iy_i$ plane.

Lenses $L_2$ and $L_3$ have the same focal length. $L_3$ projects the Fourier transform of the idler beam at the crystal on the mirror and then the Fourier transform of the field at the mirror $M_1$ back to the crystal. A phase shifter $PS$ positioned between $L_3$ and $M_1$ allows the control of the relative phases between the idler beams generated in the first and in the second passages of the pump and signal beams through the nonlinear crystal. This phase is adjusted to be $\phi$ when there is no object close to $M_2$. $L_4$ then projects the Fourier transform of the idler beams at the crystal on the detection plane. Since $L_4$ has twice the focal length of $L_2$ and $L_3$, there is a magnification by a factor of 2 \citep{goodman}. The idler field at the detector (CCD camera) is thus
\begin{eqnarray}
	E_D=&& \alpha E\left(-\frac{x}{2},\frac{y}{2};\omega_i\right)\mathrm{e}^{i\phi}+\\\nonumber
	    && +\alpha E\left(-\frac{x}{2},\frac{y}{2};\omega_i\right)\left[t\left(-\frac{x}{2},-\frac{y}{2}\right)\right]^2\mathrm{e}^{2i\gamma(-x/2,-y/2)},
\end{eqnarray}
where the first term in the right side represents the idler beam component that was generated in the first passage of the pump and signal beams through the nonlinear crystal. The second term in the right side represents the idler beam component that was generated in the second passage of the pump and signal beams through the crystal. The light intensity at the detector is 
\begin{align}\label{int}
	I(x,y)\propto&\left| E\left(-\frac{x}{2},\frac{y}{2};\omega_i\right)\right|^{2}\nonumber \\ 
	               &\times \left|1+ \left[t\left(-\frac{x}{2},-\frac{y}{2}\right)\right]^2\mathrm{e}^{i[2\gamma(-x/2,-y/2)-\phi]}\right|^2.
	               \end{align}
By detecting the intensity $I(x,y)$ on the CCD with different phases $\phi$, the object amplitude $t(x,y)$ and the phase pattern $\gamma (x,y)$ can be obtained. From Eq. (\ref{int}), it can be seen that in each detector position $(x,y)$ we will have a sinusoidal variation of the intensity with a controlled variation of $\phi$. The visibility $V(x,y)$ associated to each position is
\begin{align}\label{intvis}
	V(x,y) =&
	\frac{I_\mathrm{max}-I_\mathrm{min}}{I_\mathrm{max}+I_\mathrm{min}}\nonumber\\
	=&\frac{2\left[t\left(-\frac{x}{2},-\frac{y}{2}\right)\right]^2}{1+\left[t\left(-\frac{x}{2},-\frac{y}{2}\right)\right]^4},
\end{align}
where $I_\mathrm{max}$ and $I_\mathrm{min}$ represent the maximum and minimum intensities of the interference pattern at the position $(x,y)$, respectively. So by measuring $V(x,y)$ and inverting Eq. (\ref{intvis}) we can recover $t(x,y)$. The intensity maxima occur when $\gamma (-x/2,-y/2)=\phi/2$, so that by measuring $I_\mathrm{max}(x,y)$ for several different values of $\phi$ we can recover $\gamma (x,y)$. 
 
In the cases of ``pure amplitude'', $t(x,y)$ with $\gamma=0$ and ``pure phase'', $\gamma(x,y)$ with $t=1$  objects, an alternative method can be used. Replacing the visibility by a contrast
\begin{equation}
\label{cont}
C(x,y)=\frac{I_0}{I_0+I_\pi},
\end{equation}
where $I_0$ and $I_\pi$ refer to the intensities measured with $\phi=0$ and $\phi=\pi$, respectively, it is easy to work out Eq.~(\ref{int}) to show that 
\begin{equation}
\label{tr}
t(x,y)=\left\{\frac{1-2\sqrt{C(x,y)-[C(x,y)]^2}}{2C(x,y)-1}\right\}^\frac{1}{2},
\end{equation}
when $\gamma(x,y)\equiv 0$, and
\begin{equation}
\label{gr}
\gamma(x,y)=\frac{1}{2}\arccos[2C(x,y)-1],
\end{equation}
when $t(x,y)\equiv 1$.
 
\begin{figure}
\centering
\includegraphics{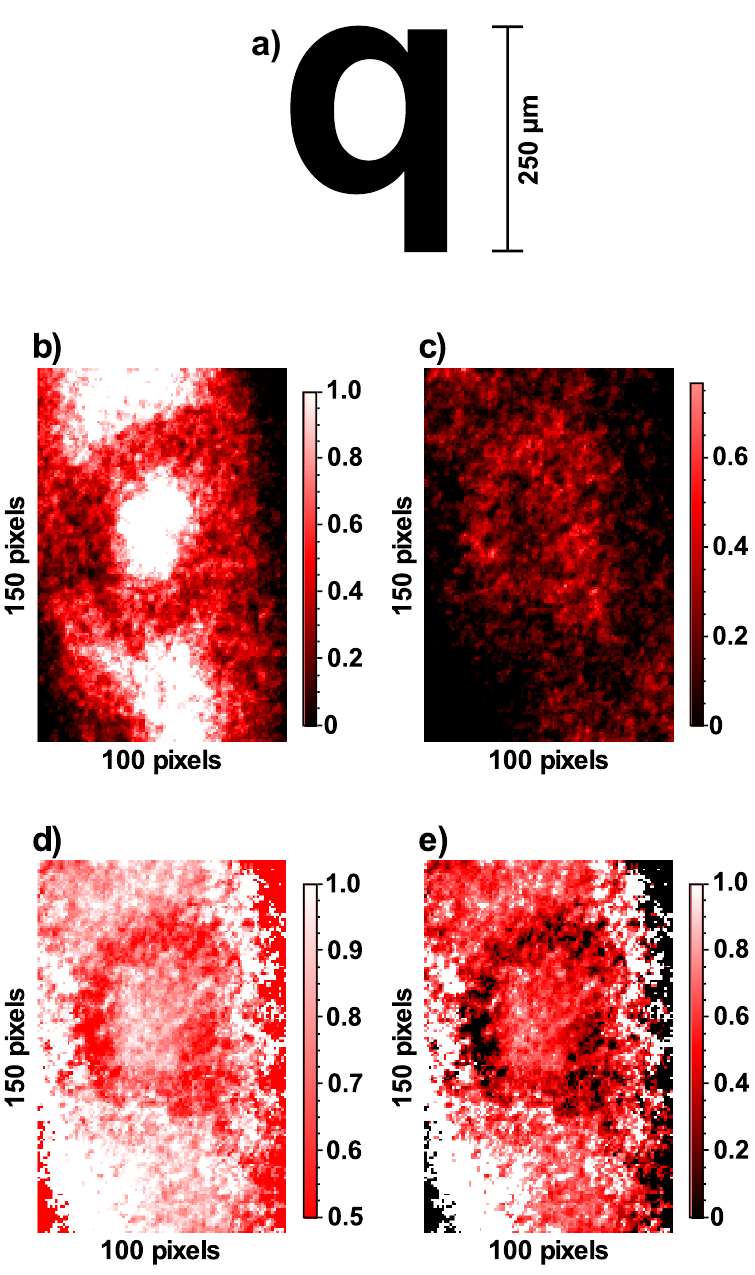}
\caption{Imaging of transmission object in pixel scale (pixel dimensions of the CCD are 6.7 $\mu$m $ \times$ 6.7 $\mu$m). (a) Letter `q' printed on a transparency film is the object positioned close to mirror $M_2$. (b) Relative intensity image on the CCD camera for a constructive interference condition, $\phi=0$ in Eq. (\ref{int}).  (c) Relative intensity image on the CCD camera for a destructive interference condition, $\phi=\pi$ in Eq. (\ref{int}). (d) Contrast calculated with Eq. (\ref{cont}). (e) Transmission pattern $t(x,y)$ recovered using Eq. (\ref{tr}). The color scale used is proportional to the relative intensity in (b) and (c), to the contrast in (d) and to the transmittance in (e).} \label{fig2}
\end{figure}

We first illustrate our method by obtaining the relative intensity image of a letter `q' printed on a transparency film positioned very close to the mirror $M_2$ in the scheme of Fig. 1. Fig. \ref{fig2}(a) shows the transmission object. Fig. \ref{fig2}(b) displays the relative intensity image on the CCD camera for $\phi=0$  and Fig. \ref{fig2}(c) displays the relative intensity image for  $\phi=\pi$. The contrast $C(x,y)$ and recovered image of $t(x,y)$ using Eq. (\ref{tr}) are displayed in Figs. \ref{fig2}(d) and \ref{fig2}(e), respectively. It should be noticed that the contrast $C(x,y)$ for a pure transmission object should lie between 0.5 and 1. However, due to intensity and pixel readout fluctuations, as well as residual phase variations in the transparency, $C(x,y)$ may assume values less than 0.5 for some pixels. This leads to imaginary values for $t(x,y)$ in Eq. (\ref{tr}). Those pixels (6.6\% total) were discarded when applying Eq. (\ref{tr}). 

\begin{figure}
\centerline{\includegraphics{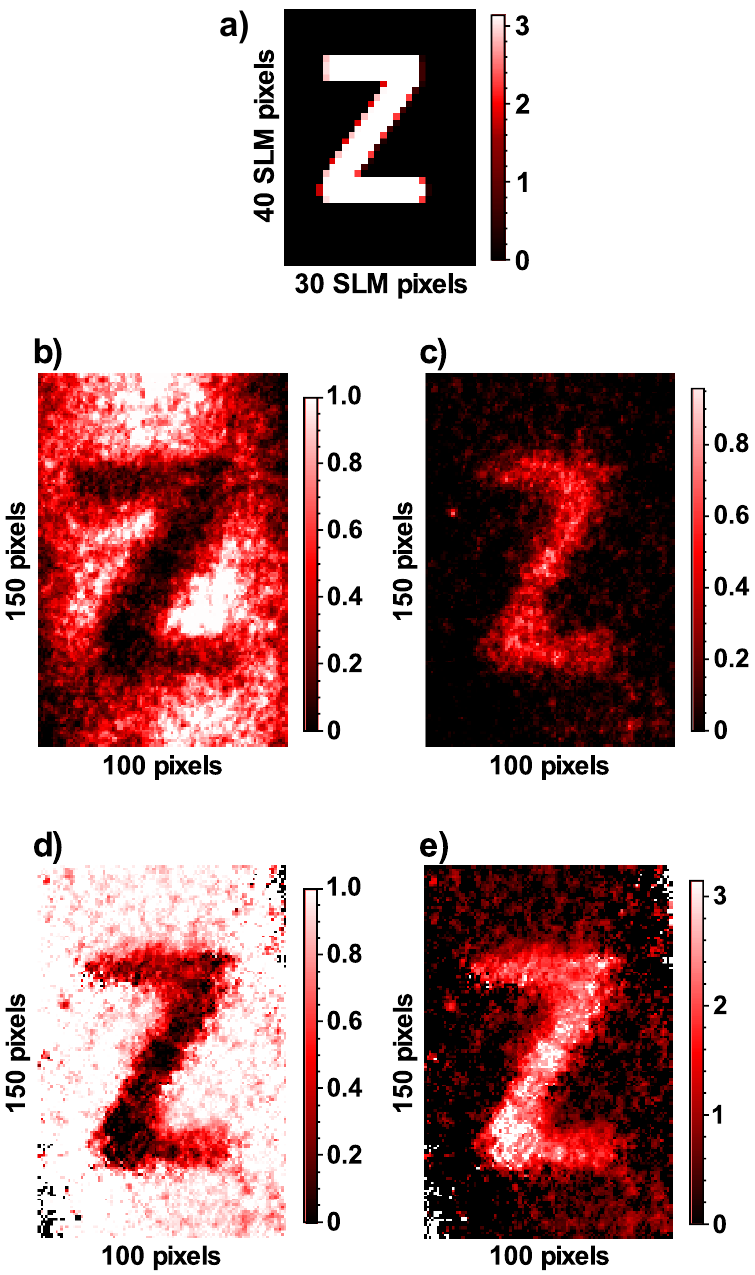}}
\caption{Imaging of a phase object. (a) Phase pattern of the phase object in pixel scale generated by a phase spatial light modulator in the position of $M_2$. Pixel dimensions of the SLM are 20 $\mu$m $\times $ 20 $\mu$m. (b) Relative intensity image in pixel scale on the CCD camera for a constructive interference condition, $\phi=0$ in Eq. (\ref{int}).  (c) Relative intensity image in pixel scale on the CCD camera for a destructive interference condition, $\phi=\pi$ in Eq. (\ref{int}). (d) Contrast calculated with Eq. (\ref{cont}).  (e) Phase pattern $\gamma(x,y)$ recovered using Eq. (\ref{gr}). The color scale used is proportional to the relative intensity in (b) and (c), to the contrast in (d) and to the phase (radians) in (e).} \label{fig3}
\end{figure}

As a second example, a phase image is produced by replacing the mirror $M2$ and the transmission object O in Fig. \ref{fig1} by a reflection type phase-only spatial light modulator (SLM). The SLM pixel size of 20 $\mu$m $\times $ 20 $\mu$m allows a good spatial resolution for the object exhibited in the liquid crystal display. 
Different gray scales on the SLM's display add different phases on the reflected light field. Fig. \ref{fig3}(a) shows the phase pattern used. Inside the letter `z' the phase is $\pi$ and outside it the phase is zero. 
This object has therefore a reflection coefficient $r(x,y)\mathrm{e}^{i\gamma(x,y)}$, with $r(x,y)$ and $\gamma(x,y)$ being real functions, in the coordinate axes given by $(x_s, y_s, z_s)$ in Fig. 1. There is not a signal double passage through this reflection object as assumed above for a transmission object and therefore Eqs. (\ref{int}) and  (\ref{intvis}) are modified by doing the following substitution: $[t(-x/2,-y/2)]^2 \to r(-x/2,-y/2) $ and $ 2 \gamma(-x/2,-y/2) \to \gamma(-x/2,-y/2) $. In an ideal reflection phase object, $ r(-x/2,-y/2) $ = 1. For our SLM, $ r(-x/2,-y/2)$ = 0.94.
The phase shifter is again adjusted for $\phi=0$ in the CCD image of Fig. \ref{fig3}(b) and $\phi=\pi$ in the image of Fig. \ref{fig3}(c). The contrast $C(x,y)$ and the recovered phase pattern $\gamma(x,y)$ using Eq. (\ref{gr}) are shown in Figs. \ref{fig3}(d) and \ref{fig3}(e), respectively.

To summarize, we were able to construct intensity and phase images of an object by detecting classical light which never interacted with it. We showed therefore that the method used in Ref. \citep{Nature}, where quantum images of objects were produced by detecting photons that did not interact with it,  has a classical analogue. As in the case of the entangled states produced in Ref. \citep{Nature}, classical states of light produced by stimulated parametric downconversion also have a high degree of spatial and phase correlations. We have used these correlations to achieve our results. The truly quantum aspect of Ref. \citep{Nature} is that high quality images were produced without stimulated emission, which cannot be explained by classical physics.

Besides the fundamental implications of this work, there are potential applications, especially because the frequency of the photons that interact with the object may be different from the frequency of the detected photons. Although the wavelength chosen for the idler and signal beams here are close, this is not necessary. Any two wavelengths that satisfy the phase matching condition can be explored. This fact can be used for optimizing the efficiency of the imaging procedure when the light that illuminates the object must have a frequency far from its absorption range. In biological samples, for instance, it may be useful to illuminate the samples with infrared radiation avoiding excessive light absorption and the consequent sample modification, while the light detectors are optimized to optical frequencies in the visible. Our technique can have the same future applications as the ones shown in Ref. \citep{Nature}, except those which require quantum fields.

It is interesting to note that no phase locking between the input pump and signal beams is necessary to construct the image with the idler beam.  In the stimulated down-conversion regime it is possible to speak about the phase correlation between the pump and the down-converted beams: the sum of the phases of the down-converted beams is equal to the phase of the pump beam.  Therefore, the phase locking is intrinsic. 
A parallel between the stimulated down-converted imaging scheme  and quantum illumination can be established. In both schemes a nonlinear crystal performs the correlation although in quantum illumination entanglement has a special role \citep{lloyd,shapiro}. In quantum illumination it is observed an effective signal-to-noise ratio gain in the detection of a weak reflecting noisy target by doing joint measurement between the signal photon transmitted by the target and the ancilla idler photon. It would be interesting to investigate this aspect also here as a classical counterpart.
 Another interesting feature is that the coherence length of the idler and signal beams can be very high, being defined by the coherence length of the input laser beams. This is not the case in the experiments discussed in reference \citep{Nature}, since the coherence length of the photons generated by spontaneous parametric downconversion is determined by the inverse of the bandwidth frequency of the interference filters used at the detectors, usually being much smaller than what we have here. Our classical imaging technique can thus be used to image objects which present large variation in thickness throughout its profile. The high gain regime of our experiment also enables the high contrast imaging of objects that have very low mean transmittance \citep{Pra270.05,2040-8986-19-5-054003,CIWUP}, which is often the case with biological samples. The length difference between the interferometer arms in our experiment do not need to be controlled with high precision as in the quantum version of Ref. \citep{Nature}, making the classical interferometer easier to implement and more robust to noise. Another advantage is that classical fields have higher intensities and can generate an image much faster than in the quantum limit, where photons are detected one by one.  
\acknowledgments
This research was supported by the Brazilian agencies CAPES, CNPq and FAPEMIG. The authors acknowledge Raphael Drumond for useful discussions and suggestions.


\begin{thebibliography}{99}
\bibitem{2040-8986-18-7-073002} M. Genovese; J. Opt. \textbf{18}, 073002 (2016).
\bibitem{1464-4266-4-3-372}L. A. Lugiato, A. Gatti, and E. Brambilla; J. Opt. B \textbf{4}, S176 (2002).
\bibitem{PhysRevA.67.033812} I. F. Santos, M. A. Sagioro, C. H. Monken and S. P\'adua; Phys. Rev. A \textbf{67}, 033812 (2003).
\bibitem{PhysRevA.77.043832} I. F. Santos, L. Neves, G. Lima, C.H. Monken and S. P\'adua; Phys. Rev. A \textbf{72}, 033802 (2005).
\bibitem{2040-8986-19-1-013001} H. Rubinsztein-Dunlop \textit{et al.}; J. Opt. \textbf{19}, 013001 (2017).
\bibitem{Nature} G. B. Lemos, V. Borish, G. D. Cole, S. Ramelow, R. Lapkiewicz, and A. Zeilinger; Nature \textbf{512}, 409 (2014).
\bibitem{PhysRevA.44.4614} L. J. Wang, X. Y. Zou, and L. Mandel; Phys. Rev. A \textbf{44}, 4614 (1991).
\bibitem{PhysRevLett.67.318} X. Y. Zou, L. J. Wang, and L. Mandel, Phys. Rev. Lett. \textbf{67}, 318 (1991).
\bibitem{PhysRevA.53.2804}T. B. Pittman, D. V. Strekalov, D. N. Klyshko, M. H. Rubin, A. V. Sergienko and Y. H. Shih; Phys. Rev. A \textbf{53}, 2804 (1996); B. Pittman, Y. H. Shih, D. V. Strekalov and A. V. Sergienko; Phys. Rev. A \textbf{52}, R3429 (1995).
\bibitem{1367-2630-15-7-073032} R. S. Aspden, D. S. Tasca, R. W. Boyd, and M. J. Padgett, N. J. Phys. \textbf{15}, 073032 (2013).
\bibitem{PhysRevLett.93.213903}A. F. Abouraddy, P. R. Stone, A. V. Sergienko, B. E. A. Saleh and M. C. Teich; Phys. Rev. Lett. \textbf{93} 213903 (2004).
\bibitem{PhysRevA.58.605} A. G. White, J. R. Mitchell, O. Nairz and P. G. Kwiat; Phys. Rev. A \textbf{58}, 605 (1998).
\bibitem{Advoptphoton.2.405} Baris I. Erkmen and Jeffrey H. Shapiro; Adv. Opt. Photon. \textbf{2}, 405 (2010).
\bibitem{PhysRevA.92.013832} M. Lahiri, R. Lapkiewicz, G. B. Lemos and A. Zeilinger; Phys. Rev. A \textbf{92}, 013832 (2015).
\bibitem{Ataman2016} S. Ataman; Eur. Phys. J. D \textbf{70}, 127 (2016).
\bibitem{2040-8986-19-5-054003} M. I. Kolobov, E. Giese, S. Lemieux, R. Fickler and
R. W. Boyd; J. Opt. \textbf{19}, 054003 (2017).
\bibitem{CCPDIL} Z. Y. Ou, L. J. Wang, X. Y. Zou and L. Mandel; Phys. Rev. A \textbf{41}, 1597 (1990).
\bibitem{OSAWang} L. J. Wang, X. Y. Zou and L. Mandel; J. Opt. Soc. Am. B \textbf{8}, 978 (1991).
\bibitem{PhysRevA.51.1631} P. H. Souto Ribeiro, S. P\'adua, J. C. Machado da Silva and G. A. Barbosa; Phys. Rev. A \textbf{51}, 1631 (1995).
\bibitem{PhysRevA.60.5074} P. H. Souto Ribeiro, S. P\'adua and C. H. Monken; Phys. Rev. A \textbf{60}, 5074 (1999).
\bibitem{PRL72} T.J. Herzog, J.G. Rarity, H. Weinfurter and A. Zeilinger; Phys. Rev. Lett. \textbf{72}, 629 (1994).
\bibitem{Pra270.05} H.M. Wiseman and K. M$\o$lmer; Phys. Rev. A \textbf{270}, 245 (2000).
\bibitem{CIWUP} J. H. Shapiro, D. Venkatraman and F. N. C. Wong; Sci. Rep. \textbf{5}, 1039 (2015).
\bibitem{yurke} B. Yurke, Phys. Rev. Lett. \textbf{56}, 1515 (1986). 
\bibitem{plick} W. N. Plick, J. P. Dowling, and G. S. Agarwal, New J. Phys. \textbf{12}, 083014 (2010).
\bibitem{hudelist} F. Hudelist, Jia Kong, Cunjin Liu, Jietai Jing, Z.Y. Ou, Weiping Zhang,  Nat. Commun. \textbf{5}, 3049 (2014).
\bibitem{boyd} R. Boyd, Nonlinear Optics (Elsevier Science, New York, 2013).
\bibitem{goodman} J. Goodman, Introduction to Fourier Optics (Roberts and Company Publishers, Greenwood Village, 2005).
\bibitem{lloyd} S. Lloyd; Science \textbf{321}, 1463 (2008).
\bibitem{shapiro}Si-Hui Tan, Baris I. Erkmen, Vittorio Giovannetti, Saikat Guha, Seth Lloyd, Lorenzo Maccone, Stefano Pirandola, and Jeffrey H. Shapiro; Phys. Rev. Lett. \textbf{101}, 253601 (2008).


\end{thebibliography}
\end{document}